\documentclass{article}[12pt]
\usepackage{epsfig}     
\usepackage{amsmath}    
\usepackage{amssymb}   
\newcommand{\beeq}{\begin{eqnarray}}     
\newcommand{\eeeq}{\end{eqnarray}}     
\newcommand{\be}{\begin{equation}}     
\newcommand{\ee}{\end{equation}}

\newcommand\npb[3]{{\it Nucl. Phys. }{\bf B #1} (#2) #3}     
\newcommand\npa[3]{{\it Nucl. Phys. }{\bf A #1} (#2) #3}     
     
\newcommand\plb[3]{{\it Phys. Lett. }{\bf B #1} (#2) #3}

\newcommand\prd[3]{{\it Phys. Rev. }{\bf D #1} (#2) #3}     
\newcommand\prl[3]{{\it Phys. Rev. Lett. }{\bf  #1} (#2) #3} 
     
\newcommand\zpc[3]{{\it Z. Phys. }{\bf C #1} (#2) #3}     
  
\def\xb{{\bf x}}     
\def\yb{{\bf y}}      
\def\zb{{\bf z}}   
\def\asb{{\bar{\alpha}_s}}
      
\begin{document}
\titlepage      
\begin{flushright}     
March 2005      
\end{flushright}      
      
\vspace*{1in}      
\begin{center}      
{\Large \bf Algebraic models for the hierarchy structure of evolution equations 
at small ${\bf x}$ }\\      
\vspace*{0.4in}      
P. Rembiesa$^{(a)}$      
and A.\ M. \ Sta\'sto$^{(b,c)}$ \\      
\vspace*{0.5cm}       
{\it $^{(a)}$ Physics Department, The Citadel, Charleston, SC 29409, USA}\\      
$^{(b)}$ {\it Nuclear Theory Group, Brookhaven National Laboratory, Upton, NY 11973, USA\footnote{Permanent address.}} \\      
$^{(c)}${\it Institute of Nuclear Physics PAN, Radzikowskiego 152,      
 Krak\'ow, Poland}
 \vskip 2mm      
\end{center}      
\vspace*{1cm}      
\centerline{(\today)}      
      
\vskip1cm      
\begin{abstract}    
We explore several  models of QCD evolution equations simplified by considering only the rapidity dependence of dipole scattering amplitudes, while provisionally neglecting their dependence on transverse coordinates.   Our main focus is on the equations that include the processes of pomeron splittings. We examine the algebraic structures of the governing equation hierarchies, as well as the asymptotic behavior of their solutions in the large-rapidity limit. 
\end{abstract} 
\newpage

\section{Introduction}

 Parton saturation and the unitarity bound  are among the interesting, yet still not fully resolved problems of the high energy QCD. There are two equivalent descriptions of  parton saturation effects:  the Balitsky hierarchy, \cite{Bal}  and the JIMWLK functional equation \cite{JIMWLK}.
 The Balitsky hierarchy consists of an infinite set of coupled evolution equations for the operators involving Wilson lines. They describe quark and gluon scattering off a target, and generally are applicable in the high-energy regime. On the other hand, the JIMWLK equation is a functional equation which describes the evolution of the target in the limit of small $x$.   
 When the JIMWLK hamiltonian acts onto a dipole operator consisting of  Wilson lines in the fundamental representation, the Balitsky hierarchy is reproduced. In the limit of negligible correlations within the target, the first equation of the Balitsky hierarchy decouples, and becomes equivalent to the nonlinear Kovchegov equation \cite{Kov}.  The latter has been derived within the dipole formalism \cite{Mueller94} by taking into the account multiple scatterings between the dipoles in the projectile wavefunction and the target.

A common feature of the above approaches is that they involve triple interactions of BFKL pomerons, \cite{Bartels} and thus describe the evolution of the system as a nonlinear process.  However, it has been recently recognized   that the Balitsky-JIMWLK equations incorporate only pomeron mergings, but not pomeron splittings \cite{IT}. In consequence, in the course of the evolution (understood as a process in the rapidity parameter) the number of pomerons can only decrease.  Therefore the above equations ignore  contributions from  fluctuations  from any intermediate pomeron loops. However, it is well known that such terms are relevant and can be substantial, therefore they should not be neglected \cite{Salam,MuSalam}.
Accordingly, considerable efforts have been lately undertaken towards a better understanding of the role of the fluctuations, \cite{Fluct_IM,Fluct_IMM} by including the missing pomeron splittings in the Balitsky-JIMWLK equations  \cite{IT,MuShoWong,LevLub_fluc,Levin_sol}.

Let us first consider the evolution equations as proposed in \cite{IT}.  They describe the evolution of the amplitudes $T^{(k)}_Y$ for the scattering of $k$ quark-antiquark dipoles off a target in the large number of colors $N_c \rightarrow \infty$ regime.  The rapidity $Y=\ln 1/x$ substitutes for the (imaginary) time parameter of the evolution.  The first two equations of the infinite hierarchy read 

\beeq
\frac{d T^{(1)}_Y(\xb,\yb)}{d Y}  & = & \asb \int  d^2 \zb \left[ {\cal M}(\xb,\yb,\zb) \, \otimes  \, T^{(1)}_Y(\xb,\yb) -   K_1(\xb,\yb,\zb)  T^{(2)}_Y(\xb,\zb;\zb, \yb) \right]  \; ,\nonumber \\
\frac{d T^{(2)}_Y(\xb_1,\yb_1;\xb_2, \yb_2) }{dY} & = &  \asb \int  d^2 \zb  \left[ {\cal M}(\xb_1,\yb_1,\zb) \, \otimes  \, T^{(2)}_Y(\xb_1,\yb_1;\xb_2,\yb_2) \; + \right.  \\
& & \left. +\; K_2(\xb_1,\yb_1,\xb_2,\yb_2,\zb)  \otimes T^{(1)}_Y(\xb_1,\yb_1 )  
  -  \; K_1(\xb_1,\yb_1,\zb)  T^{(3)}_Y(\dots)  \right] \; .  \nonumber
\label{eq:IT}
\eeeq

The arguments of the dipole amplitudes $T^{(k)}_Y$ are two-dimensional vectors $\xb$ and $\yb$ that give the location of the dipoles in the position space. The value of the strong coupling constant $\asb=\frac{\alpha_s N_c}{\pi}$ remains fixed, consistent with the leading $\ln 1/x$ approximation. The functions ${\cal M}, K_1$ and $K_2$ describe the BFKL kernel and the Pomeron vertices in the dipole picture, respectively.  Their exact forms can be found in \cite{IT}. Note that the first equation is coupled through $T^{(2)}_Y$ to the second one, the second one is coupled through $T^{(3)}_Y$ to  the third one, and so on.
 This way the infinite hierarchy is formed.
   
Describing the exact behavior of the solutions to this hierarchy poses a nontrivial task. Even in the case of the closed nonlinear Kovchegov equation, the exact analytical solution is not available, although one can gain some partial insight about its behavior from analytic properties of various approximate cases \cite{LevTu,MuPe,BaFaLi}, or numerical simulations ( see for example \cite{Weigert}). 
   
   In this contribution we study the simplified model of the infinite hierarchy (\ref{eq:IT}), in which we ignore the dependence on transverse coordinates and retain only the dependence on the evolution variable $Y$.
   This is of course a very restrictive approximation, yet nevertheless it will enable us to explore the structure of the hierarchies, the broad properties of the solutions, as well as their rapidity dependence.  In particular, we will demonstrate how to construct in a relatively simple way the general solutions to the infinite hierarchy with, and without fluctuations.  We will show that in one particular model with fluctuations included, the solutions can be written in a compact form with the differential operators expressible as Fibonacci polynomials.
We will also show that in a model that is close to the new hierarchy (\ref{eq:IT}) the  fluctuations are significant in the regime of small  amplitudes. We point out however, that in this case the physically interesting solutions do not possess a common limiting point and that they are valid only as long as $k \le 1/\alpha_s^2$, where $k$ denotes the number of dipoles.

\section{Model I without fluctuations}

Let us consider a simplified model for the infinite hierarchy of equations
\be
\frac{dx_k(t)}{dt} \; = \; x_k(t) - x_{k+1}(t) \; ,
\label{eq:bal}
\ee
where $x_k(t)$ is the amplitude for the scattering of $k$ dipoles off the target.  Here it depends only on the evolution variable $t$.
The first term on the right-hand side is responsible for the free  propagation,  while the  second term describes the process responsible for
pomeron merging. 

Let us rewrite the equation (\ref{eq:bal}) in a matrix form,
\be
\frac{d}{dt}{{\bf X}(t)} \; = \; {\bf M} \,  {\bf X}(t) \; ,
\label{eq:h1}
\ee
where 
\begin{displaymath}
{\bf X}(t) \; = \;
\left(\begin{array}{c}
x_1(t) \\
x_2(t) \\
x_3(t) \\
\cdot\\
\cdot \\
x_n(t) \\
\cdot\\
\cdot
\end{array}\right) \; 
\end{displaymath}
is the vector of the solutions for the amplitudes $x_k(t)$.
  In the present model the matrix ${\bf M}$ has the form
\begin{displaymath}
{\bf M} \; = \; 
\left(
\begin{array}{ccccccc}
1 & -1 & 0 & 0 & 0 & 0 & \dots \\
0 & 1 & -1 & 0 & 0 & 0 & \dots \\
0 & 0 & 1 & -1 & 0 & 0 & \dots \\
0 & 0 & 0 & 1 & -1 & 0 & \dots \\
0 & 0 & 0 & 0 & 1 & -1 & \dots \\
0 & 0 & 0 & 0 & 0 & 1 & \dots \\
& & & \cdot & & & \\
& & & \cdot & & & \\
\end{array}
\right) \; .
\end{displaymath}

Note that every finite matrix ${\bf M^{(n)}}$
that corresponds to the truncated version of (\ref{eq:h1}) consisting of just $n$ equations has the determinant ${\rm \bf det}{\bf M^{(n)}}=1$ and has $n$ degenerate eigenvalues $\lambda_1=\dots=\lambda_n=1$. However, in order to solve the complete set (\ref{eq:h1}), it is required to find the eigenvectors of the full infinite matrix ${\bf M}$. It is straightforward to notice that the vector
\begin{displaymath}
{\bf V} \; = \;
\left(\begin{array}{c}
1 \\

0 \\
0 \\
\cdot\\
\cdot \\
0 \\
\cdot\\
\cdot
\end{array}\right) \; ,
\end{displaymath}
solves the eigenvalue problem
\be
{\bf M} \, {\bf V} = \lambda {\bf V} \; ,
\ee
with $\lambda=1$. This defines a particular solution of the infinite hierarchy (\ref{eq:h1})
\beeq
x_1(t) & =  & e^t\, , \nonumber \\
x_k(t) & = & 0\, , \; k>1  \; ,
\eeeq
which corresponds to the free propagation of a single particle analogous to the BFKL pomeron.
To find a more general set of solutions we postulate that
\be
x_1(t) \; = \; e^t \, f(t) \; ,
\ee
where $f(t)$ represents an arbitrary function. By recursive use of the equations (\ref{eq:bal})
we find that
\be
x_{k+1}(t) \; = \;(-1)^{k} \,  e^t \,  \frac{d^{k} f }{dt^{k}}\, , \; k \ge 0 \; ,
\label{eq:solgen1}
\ee
where $\frac{d^0 f }{dt^0} \equiv f$. So far, the only restriction on the function $f(t)$ is that it belongs to  $C^{\infty}$ .
These solutions can be given a more compact form with help of a generating functional ${\cal G}$ for  $x_k$
\be
{\cal G}[u]  \; \equiv \;  \sum_{k=0}^\infty \; u^k \, x_{k+1}  \; ,
\label{eq:gen1}
\ee
or exactly,
\be
{\cal G}[u] \; = \; e^t \frac{1}{1+u\, \frac{d}{dt}} \, f(t) \; ,
\ee
where it is understood that the differential operator acts only on the function $f(t)$ which in fact determines the specific solution. For example, the choice of $f(t) = e^{-t}$  corresponds to constant solutions
 \be
 x_1=x_2=\dots =x_k=\dots = 1 \; .
 \label{eq:solpar1}
 \ee  
 
 In order to find a nontrivial solution, let us provisionally assume that
 \be
 x_2 \; = \; \gamma x_1 \, x_1, 
 \label{eq:fact}
 \ee
where $\gamma$ is a non-zero constant. In the case of $\gamma \neq 1$
, it can be interpreted as a source of non-zero correlations \cite{LevLub_corr,Janik}  .
From the assumption (\ref{eq:fact})  it follows that the function $f(t)$ satisfies the differential equation
\be
\frac{df}{dt} = - \gamma f^2 e^t\; ,
\ee
whose solution is
\be
f(t) \, = \, \frac{1}{\gamma}\frac{1}{e^t+C}\; .
\ee
Using this solution in (\ref{eq:solgen1}) we readily find that
\beeq
x_1(t) & = &  \frac{1}{\gamma}\frac{e^t}{e^t+C} \; ,\nonumber \\
x_2(t) & = &  \frac{1}{\gamma}\left(\frac{e^t}{e^t+C}\right)^2 \; ,\nonumber \\
x_3(t) & = &  \frac{1}{\gamma} \left[\frac{2\,e^{3\,t}}{{\left( e^t  + C \right) }^3} - 
    \frac{e^{2t}}{{\left( e^t + C\right) }^2}\right] \; ,\\
x_4(t) & = &  \frac{1}{\gamma} \left[  \frac{6\,e^{4\,t}}{{\left(  e^t  + C\right) }^4} -
  \frac{6\,e^{3\,t}}{{\left(  e^t + C\right ) }^3} + 
  \frac{e^{2t}}{{\left( e^t  + C\right) }^2}\right] \; ,\nonumber \\
& \cdot & \nonumber \\
& \cdot & \nonumber 
\label{eq:solpar2}
\eeeq
Interestingly, in the  $t \rightarrow \infty$ limit, all the amplitudes $x_k \rightarrow 1/\gamma$. In order to satisfy the unitarity bound for the amplitudes, it is required that $x_k \le 1$ and, accordingly, $\gamma \ge 1$.

\section{Model II for  the hierarchy with fluctuations}

Let us now modify the hierarchy by enriching it with terms responsible
for particle splitting \cite{IT,MuShoWong,LevLub_fluc}
\be
\frac{dx_k(t)}{dt} \; = \; x_k(t) - x_{k+1}(t) +x_{k-1}(t)\, , \; k \ge 1,
\label{eq:balf}
\ee
(with $x_0 \equiv 0$).
Again,
\be
\frac{d}{dt}{{\bf X}(t)} \; = \; {\bf M} \,  {\bf X}(t) \; ,
\label{eq:h2}
\ee
where now
\begin{displaymath}
{\bf M} \; = \; 
\left(
\begin{array}{ccccccc}
1 & -1 & 0 & 0 & 0 & 0 & \dots \\
1 & 1 & -1 & 0 & 0 & 0 & \dots \\
0 & 1 & 1 & -1 & 0 & 0 & \dots \\
0 & 0 & 1 & 1 & -1 & 0 & \dots \\
0 & 0 & 0 & 1 & 1 & -1 & \dots \\
0 & 0 & 0 & 0 & 1 & 1 & \dots \\
& & & \cdot & & & \\
& & & \cdot & & & \\
\end{array}
\right)\; ,
\end{displaymath}
or symbolically,
\be
{\bf M} \; = \; {\bf 1} \, - \, {\bf A} \, + \,  {\bf A^{T}},
\ee
where $A_{ij}=\delta_{i\, j-1}$. 
The elements of ${\bf A}$ are responsible for  transitions from $k$ to  $k-1$ particles,
while the elements of ${\bf A^T}$ generate transitions from $k-1$ to $ k$ particles.
It is straightforward to construct a generalization to a more complicated structure with $k$-to-$n$ particle transitions with arbitrary $k,n$ once the corresponding vertices are known.

The problem now becomes somewhat less degenerate since  the finite $(n\times n)$ submatrices ${\bf M^{(n)}}$ of the truncated hierarchy have different eigenvalues.
We also remark that in the present case the determinant ${\rm \bf det}\; {\bf M}$ is infinite.

The fact that the matrix ${\bf M}$ has only a finite number of non-zero terms in any given row or column makes it easy to construct an eigenvector.  For example, the eigenvector to the eigenvalue $\lambda=1$ of the problem
\be
{\bf M } {\bf W } = \lambda {\bf W} \; ,
\ee
is
\begin{displaymath}
{\bf W} \; = \;
\left(\begin{array}{c}
1 \\
0 \\
1 \\
0\\
1 \\
0 \\
\cdot\\
\cdot
\end{array}\right)\; .
\end{displaymath}
We note that this vector is not normalizable; ${\bf W}^{\dagger}{\bf W}=\infty$.
Nevertheless, we shall adopt it as a  particular solution of (\ref{eq:h2}), whose exact form is
\beeq
x_{2l-1}(t)  & = & e^t \, ,\nonumber \\
x_{2l}(t) & = & 0, \; l \ge 1 \; .
\eeeq
Again, one can construct a generalized solution by taking an arbitrary function $f(t)$, postulating that $x_1(t) = e^t f(t)$, and using the hierarchy equations to generate the remaining $x_k(t)$.  As before, the specific form of $f(t)$ is dictated by the type of the solutions sought.
In the general case, the first six amplitudes are
\beeq
x_1(t) & = & e^t f \; , \nonumber \\
x_2(t) & = & -e^t f^{(1)}  \; ,\nonumber \\
x_3(t) & = & e^t \left(f+f^{(2)}\right) \; , \nonumber \\
x_4(t) & = & -e^t \left(2 f^{(1)} + f^{(3)}\right)  \; , \\
x_5(t) & = & e^t \left(f +3 f^{(2)} + f^{(4)} \right) \;,  \nonumber \\
x_6(t) & = & -e^t \left(3f^{(1)} +4 f^{(3)} + f^{(5)} \right)  \; , \nonumber
\eeeq
where  $f^{(k)} = \frac{d^k f}{dt^k}$.

The perusal of the coefficients of the derivatives $f^{(k)}$ reveals that they are the same as in the Fibonacci polynomials, therefore we can rewrite the solution in a more compact form,
\be
x_{k+1}(t) \; = \; (-1)^{k} \, e^t \, {\cal F}_{k}[\frac{d}{dt}] \, f(t) , \; k \ge 0,
\label{eq:solpar4}
\ee
where  ${\cal F}_{k}[u]$ are the Fibonacci polynomials (for the case of  ${\cal F}_{0}[u]=1$).
The generating functional does not become much more complicated, and it reads
\be
{\cal G}[u] \;  =  \; e^t \frac{1}{1+u \frac{d}{dt} -u^2} f(t) \; .
\ee

If we again assume the approximate factorization (\ref{eq:fact}) then 
the corresponding particular solution is
\beeq
x_1(t) & = &  \frac{1}{\gamma}\frac{e^t}{e^t+C} \; ,\nonumber \\
x_2(t) & = &  \frac{1}{\gamma}\left(\frac{e^t}{e^t+C}\right)^2 \; ,\nonumber \\
x_3(t) & = & \frac{1}{\gamma} \left[ 2\left(\frac{e^t}{e^t+C} \right)^3-\left(\frac{e^t}{e^t+C} \right)^2+ \frac{e^t}{e^t+C}  \right]  \; ,\\
x_4(t) & = & \frac{1}{\gamma} \left[6\left(\frac{e^t}{e^t+C} \right)^4-6 \left(\frac{e^t}{e^t+C} \right)^3+3 \left(\frac{e^t}{e^t+C} \right)^2   \right] \; ,\nonumber \\
x_5(t) & = &  \frac{1}{\gamma} \left[ 24 \left(\frac{e^t}{e^t+C} \right)^5 - 36 \left(\frac{e^t}{e^t+C} \right)^4 + 20 \left(\frac{e^t}{e^t+C} \right)^3  - 4 \left(\frac{e^t}{e^t+C} \right)^2 + \frac{e^t}{e^t+C} \right]  . \nonumber
\label{eq:solpar5}
\eeeq
Unlike in the previous hierarchy, the  solutions $x_k$  do not approach the common 
$1/\gamma$ limit at $t\rightarrow \infty$. Instead, (for simplicity we take $\gamma=1$ )
the solution is
\beeq
x_1(\infty) & = & x_2(\infty)  =   1 \; , \nonumber \\
x_3(\infty) & =   & x_2(\infty) + x_1(\infty)  =  2 \; ,\nonumber \\
x_4(\infty) & =   & x_3(\infty) + x_2(\infty)  =  3 \; , \\
x_5(\infty) & =   & x_4(\infty) + x_3(\infty)  =  5 \; ,\nonumber \\
x_k(\infty) & = & x_{k-1}(\infty) + x_{k-2}(\infty) \, . \nonumber
\eeeq
which we recognize to be the  Fibonacci series. One of the properties of the Fibonacci series is that it exhibits the exponential behavior,
\be
F_n \; = \; [ \frac{1}{\sqrt{5}} \alpha^n ]_{\rm int},
\label{eq:Fibonacci}
\ee
where $\alpha\simeq 1.6$ and $[\dots]_{\rm int}$ stands for the nearest integer value. 
In short, in this simple model, when the amplitudes $x_1$ and $x_2$  reach the  limit 
$$x_1,x_2 = 1 \; ,$$ 
 then the  dipole amplitudes of higher-order in $k$ must exceed this limit.  However, in order to be physically meaningful, the spectrum of solutions should be bounded,
\be
0 \le x_k \le 1 \, , \; \forall k \ge 1 \; ,
 \ee
 which generally imposes an implicit condition on the behavior of $f(t)$ in (\ref{eq:solpar4}). \\
 Unfortunately, this inconsistency is fatal.  The condition of boundedness of $x_k$ requires that the derivatives $\frac{dx_k}{dt}$ vanish as $t \rightarrow \infty$ .  In this limit the hierarchy simplifies to
\be
x_{k+1} \; = \; x_k  +x_{k-1}, 
\label{eq:balfin} 
\ee
which, irrespectively of the form of the function $f(t)$, is precisely the  diverging Fibonacci series (\ref{eq:Fibonacci}).

\section{Model III for the Balitsky hierarchy}

\begin{figure}
\centering{
\includegraphics[width=0.489\textwidth]{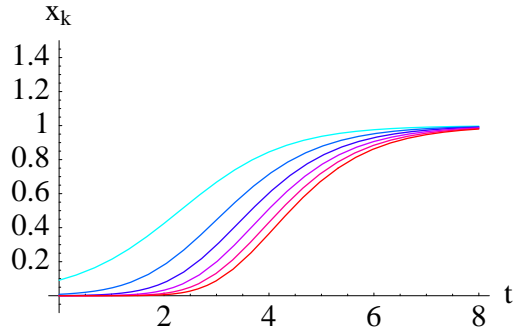}}
\caption{The factorizable solutions to the toy model for the Balitsky hierarchy (\ref{eq:balk}).
From left to right: $x_1,\dots,x_6$. The constant of integration has been set at $C=10$, and the coefficient $\gamma=1$. }
\label{fig:1}
\end{figure}

In this section we consider the model for the simplified Balitsky hierarchy \cite{Bal}
\be
\frac{dx_k(t)}{dt} \; = \;k \; \left[ x_k(t) - x_{k+1}(t) \right] \; ,
\label{eq:balk}
\ee
(compare with the equation (\ref{eq:bal})).
In this case the matrix ${\bf M}$ is still triangular, although the corresponding determinant is  infinite.  For the
truncated hierarchy it behaves as ${\bf \rm det} \, {\bf M^{(n)}}=n!$.
The eigenvector corresponding  to the eigenvalue $\lambda=1$ is the same as in the former case,
\begin{displaymath}
{\bf V} \; = \;
\left(\begin{array}{c}
1 \\
0 \\
0 \\
\cdot\\
\cdot \\
0 \\
\cdot\\
\cdot
\end{array}\right) \; .
\end{displaymath}
Therefore we can reaply the procedure developed in the last sections to obtain
\beeq
x_1(t) & = & e^t \, f^{(0)} \; ,\nonumber \\
x_2(t) & = & -e^t \, f^{(1)} \; ,  \\
x_3(t) & = & e^t \,  \left(\frac{1}{2} \left( f^{(1)}+ f^{(2)}\right)- f^{(1)}\right) \; , \nonumber \\
x_4(t) & = &  e^t \left(- f^{(1)}+\frac{1}{2} \left( f^{(1)}+ f^{(2)}\right)+\frac{1}{3}
   \left( f^{(1)}+ f^{(2)}+\frac{1}{2} \left(- f^{(1)}-2  f^{(2)}-
   f^{(3)}(t)\right)\right)\right)  . \nonumber
\eeeq
The factorization ansatz $x_2 = \gamma x_1 x_1$ leads to 
\beeq
x_k(t) =  \frac{1}{\gamma} \, \left( \frac{e^t}{e^t+C} \right)^k \; ,
\label{eq:solparf}
\eeeq
which all converge to the the same limiting point $x_k(\infty)=1$ when $\gamma=1$.
We recognize that the above are indeed the mean-field solutions of the model found in \cite{Janik,LevLub_corr}.
The first six solutions of this sequence are illustrated in Fig.~\ref{fig:1} .
\section{Model IV for the Balitsky hierarchy with fluctuations}
Let us proceed to the next model in which we add also the fluctuations terms\footnote{The solutions to the hierarchy with Pomeron loops have been recently investigated in \cite{Levin_sol}.}
to  (\ref{eq:balk})
\be
\frac{dx_k(t)}{dt} \; = \;k \; \left[ x_k(t) - x_{k+1}(t) \right] \;  +  \alpha_s^2 \, k \, (k-1) \; x_{k-1}(t)\; .
\label{eq:balkf}
\ee
The form of the coefficient $\alpha_s^2 k (k-1)$ is motivated by the structure of (\ref{eq:IT}), see \cite{IT} for details. 
In this case the matrix elements are
\be
M_{m n } =  m \delta_{mn}  - (n-1) \delta_{m \, n-1}  + \alpha_s^2 m (m-1) \delta_{m-1 \, n} \; .
\ee

The recurrence relation for the static solution $\frac{dx_k^{(s)}}{dt}=0$ is 
\be
x_{k+1}^{\rm (s)} \; = \;  x_k^{\rm (s)} + \alpha_s^2 (k-1) x_{k-1}^{\rm (s)} \; ,
\label{eq:rec_stat_IT}
\ee

which for large values of $k$  leads to 
\be
x_{k}^{(s)} \sim (\alpha_s \sqrt{k})^k \; .
\ee

This is even stronger divergence than in the  case of the Fibonacci series.  As long as $k$ is not too large, $k \ll 1/\alpha_s^2$,  the magnitudes of these static solutions remain close to each other, but for $k's$ larger than that, they quickly diverge.

\begin{figure}
\centering{
\includegraphics[width=0.489\textwidth]{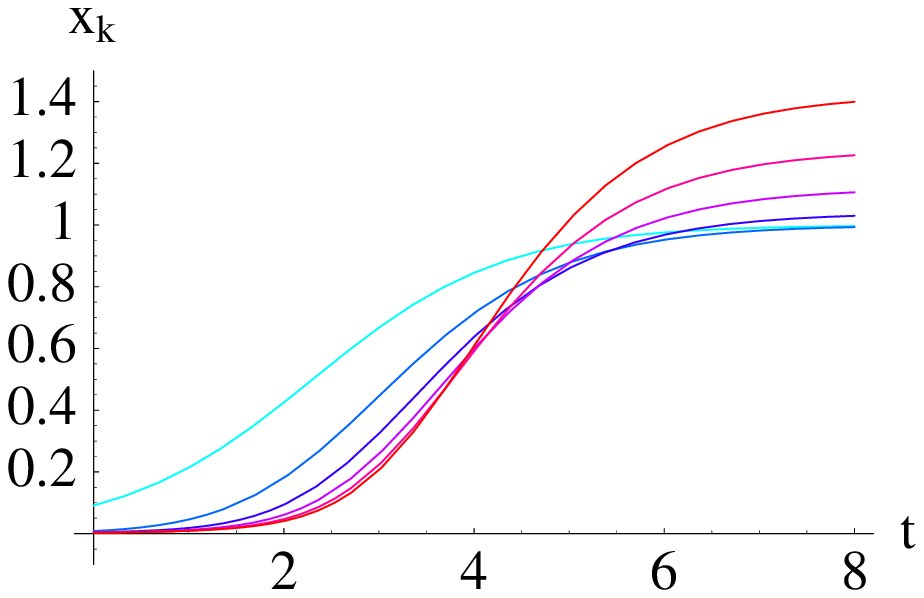}\hfill 
\includegraphics[width=0.489\textwidth]{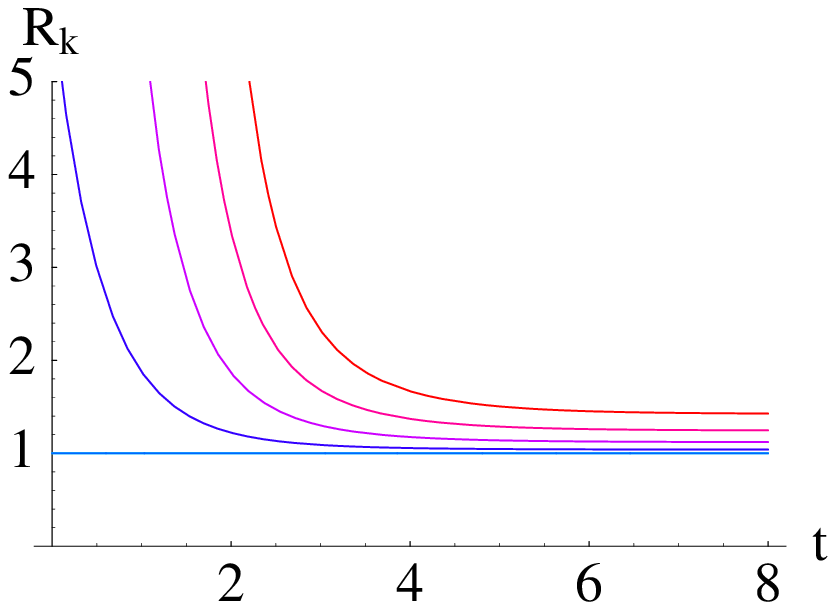}
}
\caption{Left plot: the first six solutions to the toy model hierarchy  (\ref{eq:balkf}).
The curves from the left to the right: $x_1,\dots,x_6$.
The coupling has been set to $\alpha_s = 0.2$, and the coefficient $\gamma=1$.
The constant of integration wass set at $C=10$. 
Right: The ratio of the solutions    (\ref{eq:solparfluc}) to the factorized solutions of the Balitsky hierarchy; $R_k \equiv x_k(t)/x_k^{(f)}(t)$. The constant line at $R_k=1$ is the ratio for $x_1/x_1^f$ and  $x_2/x_2^f$.
The four curves from bottom to top are $x_3/x_3^{(f)},x_4/x_4^{(f)},x_5/x_5^{(f)}$ and $x_6/x_6^{(f)}$.}
\label{fig:2}
\end{figure}

One can solve the recurrence relation (\ref{eq:rec_stat_IT}) by means of generating functional which in this case  we define as
\be
{\cal G}[u] = \sum_{n=1} u^n \frac{x_{n}^{(s)}}{n!} \; ,
\ee
(compare the definition (\ref{eq:gen1})). 
The differential equation for the functional reads
$$
\frac{d^2 {\cal G}}{du^2} = \frac{d {\cal G}}{du} (1+ \alpha_s^2 u) \; ,
$$
and has the solution
$$
{\cal G}[u] \; = \; \int_0^u  dy \,  e^{y+ \alpha_s^2 y^2/2} \; ,
$$
which can be expressed in terms of a confluent hypergeometric function $_1 F_1$
\be
{\cal G}[u] \; = \;  e^{-\frac{1}{2\alpha_s^2}} \, \frac{1+\alpha_s^2 u}{\alpha_s^2} \, \left[
_1F_1\left(\frac{1}{2},\frac{3}{2},\frac{(1+\alpha_s^2 u)^2}{2\alpha_s^2}\right) \; -
\; _1F_1\left(\frac{1}{2},\frac{3}{2},\frac{1}{2\alpha_s^2}\right)
\right]\; .
\ee
The physically interesting solutions need to be positive-definite functions. We find that the solution $x_1(t)$ from (\ref{eq:solpar5}) gives 
positive definite solutions $x_k(t)$ also in the case of the hierarchy (\ref{eq:balkf}).
The first four solutions are
\beeq
x_1(t) & = & \frac{1}{\gamma} \frac{e^t}{e^t +C} \; ,\nonumber \\
x_2(t) & = & \frac{1}{\gamma} \left( \frac{e^t}{e^t +C} \right)^2 \; ,  \\
x_3(t) & = & \frac{1}{\gamma} \left( \frac{e^t}{e^t +C} \right)^3 \; + \; \frac{\alpha_s^2}{\gamma} \frac{e^t}{e^t +C}  \; , \nonumber \\
x_4(t) & = &\frac{1}{\gamma} \left( \frac{e^t}{e^t +C} \right)^4 \; + \; \frac{\alpha_s^2}{3 \gamma} \,  \frac{e^t(9 e^t +2 C)}{( e^t +C)^2} \; . \nonumber
\label{eq:solparfluc}
\eeeq
The general form of the solutions is 
\be
x_k(t) \; = \; x_k^{(f)}(t) + \Delta x_k^{(n)}(t)  \; ,
\ee
where $x_k^{(f)}(t)$ are factorizable solutions 
$$
x_k^{(f)}(t) \; = \; (x_1(t))^k \; ,
$$
given by (\ref{eq:solparf}).
The corrections $\Delta x_k^{(n)}(t)$ are nonfactorizable and suppressed by  a factor of at least   $\alpha_s^2$. They do not vanish at $t\rightarrow \infty$, and contribute non-negligible corrections to the magnitudes of the limiting points. 
We illustrate these particular solutions (\ref{eq:solparfluc}) on the  right plot in Fig.~\ref{fig:2}, where the  differences between the positions of the limiting points $x_k(\infty)$ are apparent.
It is interesting to look in better detail at the magnitudes of the new nonfactorizable terms. For this purpose, in the right plot in Fig.~\ref{fig:2}, we show the ratio of the solutions $x_k/x_k^{(f)}$ as a function of time $t$ for the first six solutions. The solid line at $1$ corresponds to the constant ratios of $x_1/x_1^{(f)}$ and $x_2/x_2^{(f)}$. Beginning from $x_3$, the non-factorizable terms are nonzero, and they are most important at very low values of $t$, where the amplitudes $x_k$ are small, as expected \cite{IT,Fluct_IMM}.
\begin{figure}
\centering{
\includegraphics[width=0.489\textwidth]{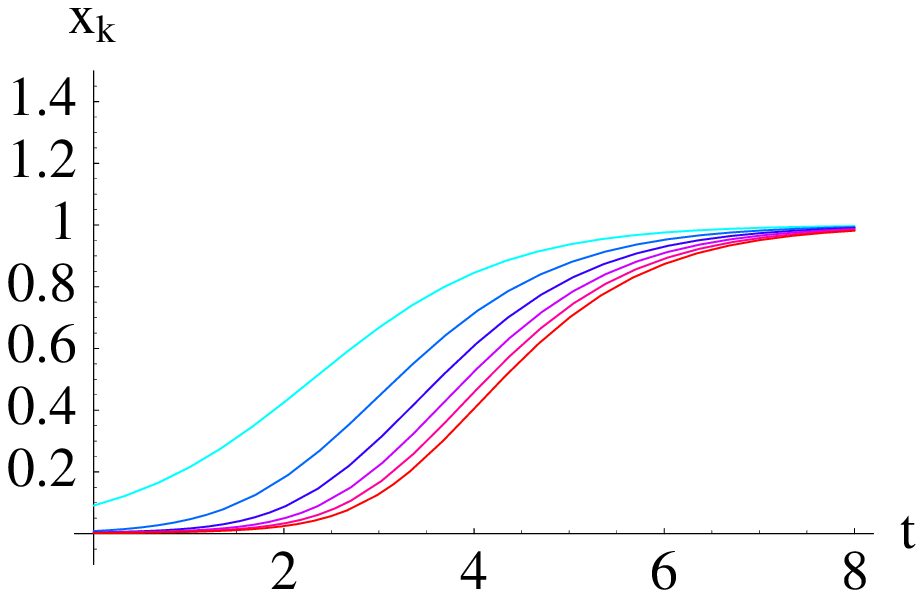}\hfill 
\includegraphics[width=0.489\textwidth]{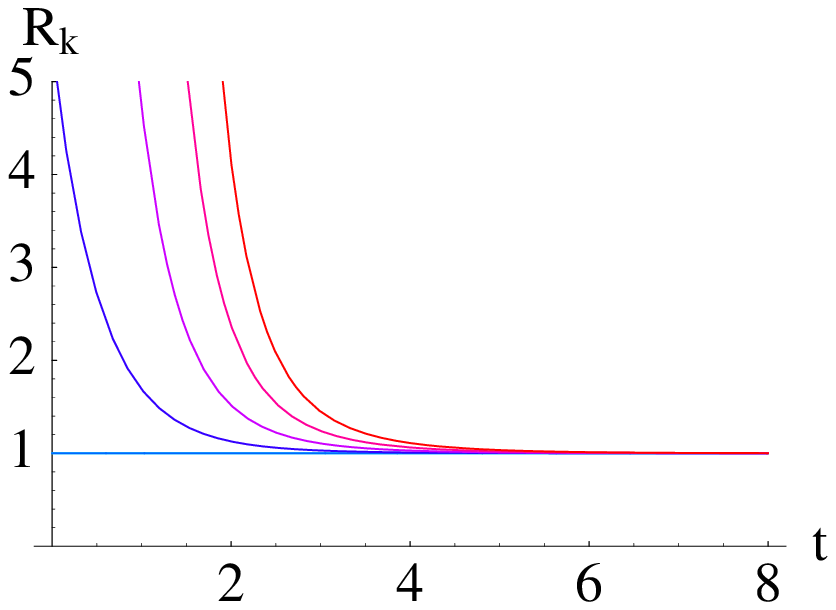}
}
\caption{Left: the first six solutions to the toy model hierarchy  (\ref{eq:balkf_ll}). The curves from the left to the right are $x_1,\dots,x_6$.
The coupling has been set to $\alpha_s = 0.2$ and the coefficient $\gamma=1$.
The constant of integration set at $C=10$.
Right: The ratio of the solutions   (\ref{eq:solparfluc_ll}) to the factorized solutions of the Balitsky hierarchy; $R_k \equiv x_k(t)/x_k^{(f)}(t)$. The constant line at $R_k=1$ is the ratio for $x_1/x_1^f$ and  $x_2/x_2^f$.
The four curves from bottom to top are $x_3/x_3^{(f)},x_4/x_4^{(f)},x_5/x_5^{(f)}$ and $x_6/x_6^{(f)}$.
}
\label{fig:3}
\end{figure}
We also consider an alternative model with an additional negative term proportional to $x_k$,
\be
\frac{dx_k(t)}{dt} \; = \;k \; \left[ x_k(t) - x_{k+1}(t) \right]  +  \alpha_s^2 \, k \, (k-1) \, \left[ x_{k-1}(t) - x_k(t) \right] \; .
\label{eq:balkf_ll}
\ee
This particular form of the hierarchy is inspired by \cite{LevLub_fluc}.
The  role of the additional term is to secure the conservation of  probability in the pomeron splitting processes.  We must  however remark that this term is formally proportional to the BFKL kernel generated by the not yet available next-to-next-to-leading order logarithmic approximation.
With the negative term included, the static solution to (\ref{eq:balkf_ll}) reads
\be
x_{k+1}^{\rm (s)} \; = \;  x_k^{\rm (s)} \; + \; \alpha_s^2 \, (k-1) \, (x_{k-1}^{\rm (s)}-x_k^{\rm (s)}) \; ,
\ee
and  has a nice property that 
\be
x_1^{(s)} =\dots=x_k^{(s)}=\dots = const \; ,
\label{eq:static_ll}
\ee
i.e. it predicts the same limiting point for all  particular solutions $x_k$, just like in the model (\ref{eq:balk}).
Now the particular solutions that satisfy the condition $x_2=\gamma x_1 x_1$  have the form
\beeq
x_1(t) & = & \frac{1}{\gamma} \frac{e^t}{e^t +C}\; , \nonumber \\
x_2(t) & = & \frac{1}{\gamma} \left( \frac{e^t}{e^t +C} \right)^2\; ,  \\
x_3(t) & = & \frac{1}{\gamma} \left( \frac{e^t}{e^t +C} \right)^3 \; + \;  
\frac{\alpha_s^2}{\gamma} \, \frac{Ce^t}{(e^t+C)^2}\; ,
  \nonumber \\
x_4(t) & = &\frac{1}{\gamma} \left( \frac{e^t}{e^t +C} \right)^4 \; + \;
\frac{\alpha_s^2}{\gamma} \frac{2e^t C(C^2+6 C e^t + 5 e^{2t})}{3(e^t+C)^4}
\;-\; \frac{\alpha_s^4}{\gamma}\frac{2 C e^t}{(e^t+C)^2} \; . \nonumber
\label{eq:solparfluc_ll}
\eeeq
As  before, we can express the solutions in the general form $x_k(t)=x_k^{(f)}(t) + \Delta x_k^{(n)}(t)$.
In this case the nonfactorizable terms vanish at least as $e^{-t}$ for $t\rightarrow \infty$,
which guarantees the convergence to the same static solution  (\ref{eq:static_ll}).
This behavior is illustrated in Fig.~\ref{fig:3}. The left plot in Fig.~\ref{fig:3} shows the first six solutions
(\ref{eq:solparfluc_ll}). The right plot shows the ratio of the solutions $x_k/x_k^{(f)}$, which  behaves similarily as in
 the case of the hierarchy (\ref{eq:balkf}). The aditional terms $\sim \alpha_s^2$ responsible for the pomeron loops are most significant in the low $t$ regime, where $x_k \ll 1$. \\

\section*{Conclusions}

We have shown that the general features of the asymptotic solutions of the hierarchies for the evolution equations in QCD can be deduced from simplified models that keep only the rapidity dependence while neglect the dependence on the transverse degrees of freedom.  We presented the general method for finding the solutions to these types of hierarchies. It is generally expected that on the grounds of the unitarity or 'black disc' conditions the solutions ought to tend to a common limit.  However we have found that in some cases of hierarchies with fluctuations the physically relevant solutions may not tend to one universal asymptotic solution. In some models they may even form highly divergent series.
We have shown that the fluctuations have the largest impact in the region of small values of amplitudes, irrespectively of the model considered.

\section*{Acknowledgments}
We thank Jochen Bartels, Krzysztof Golec-Biernat, Misha Lublinsky, Larry McLerran and Al Mueller
for discussions.
This research was supported by The Citadel Foundation, the U.S. Department of Energy, Contract No. DE-AC02-98-CH10886 and by the Polish Committee for Scientific Research, KBN Grant No. 1 P03B 028 28.


\end{document}